\documentclass[letterpaper,conference,onecolumn,12pt]{ieeeconf}
\IEEEoverridecommandlockouts                              
\usepackage{srcltx}
\usepackage{color}
\usepackage{graphicx}
\usepackage{bm}
\usepackage[cmex10]{amsmath}
\usepackage{algorithmic}
\usepackage{algorithm}
\usepackage{amssymb}
\usepackage{subfigure}
\usepackage{multirow}
\newtheorem{theorem}{Theorem}

\newtheorem{remark}{Remark}
\newtheorem{proposition}{Proposition}
\newtheorem{definition}{Definition}

\newtheorem{assumption}{Assumption}

 \newtheorem{pf}{Proof}
\usepackage{amsfonts}

\renewcommand{\Re}{\mathbb{R}}
\newcommand{\Ball}{\mathbb{B}}
\newcommand{\Ze}{\mathbb{Z}}

\newcommand{\kl}{\mathcal{KL}}

\newcommand{\ki}{\mathcal{K}_{\infty}}

\newcommand{\imp}{\rightarrow}

\newcommand{\X}{\mathcal{X}}
\newcommand{\U}{\mathcal{U}}

 \renewcommand{\baselinestretch}{1.5}
\begin{document}

\title{\bf Extremum Seeking-based Iterative Learning Model Predictive Control (ESILC-MPC)}
\author{Anantharaman Subbaraman, Mouhacine Benosman
\thanks{A. Subbaraman (anantharaman@umail.ucsb.edu) is with
the Department of Electrical Engineering, University of California
at Santa Barbara, USA. M. Benosman (m{\_}benosman@ieee.org) is
with Mitsubishi Electric Research Laboratories (MERL), Cambridge,
MA 02139, USA. }}

\maketitle

 \maketitle
\begin{abstract}
In this paper, we study a tracking control problem for linear
time-invariant systems, with model parametric uncertainties, under
input and states constraints. We apply the idea of modular design
introduced in \cite{B014}, to solve this problem in the model
predictive control (MPC) framework. We propose to design an MPC
with input-to-state stability (ISS) guarantee, and complement it
with an extremum seeking (ES) algorithm to iteratively learn the
model uncertainties. The obtained MPC algorithms can be classified
as iterative learning control (ILC)-MPC.
\end{abstract}
\section{Introduction}
Model predictive control (MPC), e.g., \cite{Mayne2000}, is a
model-based framework for optimal control of
 constrained multi-variable systems. MPC is based on the repeated,
receding horizon solution of a finite-time optimal control problem
formulated from the system dynamics, constraints on system states,
inputs, outputs, and a cost function describing the control
objective. However, since MPC is a model-based controller, its
performance inevitably depends on the quality of the prediction
model used in the optimal control computation.

In contrast, extremum seeking (ES) control is a well known
approach where the extremum of a cost function associated with a
given process performance (under some conditions) is found without
the need for detailed modelling information, see, e.g.,
\cite{ariyur2003real,AK02,Nes09}. Several ES algorithms (and
associated stability analysis) have been proposed,
\cite{k00,AK02,TNM06,Nes09,TNM06,ariyur2003real,rotea2000analysis,GDD13},
and many applications of ES have been reported
\cite{ZGD03,HGPD08,ZO12,benosmanatincacc13,benosmanatincecc13}.

The idea that we want to theoretically analyze in this paper, is
that the performance of a model-based MPC controller can be
combined with the robustness of a model-free ES learning algorithm
for simultaneous identification and control of linear
time-invariant systems with structural uncertainties. We refer the
reader to \cite{B014,benosmanatincecc13,benosmanatinccdc13} where
this idea of learning-based modular adaptive control has been
introduced in a more general setting of nonlinear dynamics

We aim at proposing an alternative approach to realize an
iterative learning-based adaptive MPC. We introduce an approach
for an ES-based iterative learning MPC that merges a model-based
linear MPC algorithm with a model-free ES algorithm to realize an
iterative learning MPC that adapts to structured model
uncertainties. Due to the iterative nature of the learning model
improvement, we want here to compare the proposed approach to some
existing Iterative learning control (ILC)-MPC methods. Indeed, ILC
method introduced in \cite{arimoto1990robustness} is a control
 technique which focuses on  improving  tracking performance of processes
 that repeatedly execute the same operation over time. It is of particular importance
 in robotics and in chemical process control of batch processes. We refer the reader
 to   \cite{wang2009survey}, \cite{moore1999iterative} and  \cite{ahn2007iterative} for
 more details on ILC and its applications.

 At the intersection of learning based control and constrained
control is the ILC-MPC concept. For instance, ILC-MPC for chemical
batch processes are studied in \cite{wang2008iterative},
\cite{cueli2008iterative}, and \cite{shi2007single}. As noted in
\cite{cueli2008iterative} one of the shortcomings of the current
literature is a rigorous justification of feasibility, and
Lyapunov-based stability analysis for ILC-MPC . For example, in
\cite{wang2008iterative} the goal is to reduce the error between
the reference and the output over multiple trials while satisfying
only input constraints. However, the reference signals is
arbitrary and the MPC scheme for tracking such signals is not
rigorously justified. Furthermore, the MPC problem does not have
any stabilizing conditions (terminal cost or terminal constraint
set). The ILC update law is an addition of the MPC signal of the
current trial to the MPC signal of the previous trial. In
\cite{cueli2008iterative}, an ILC-MPC scheme for a general class
of nonlinear systems with disturbances is proposed. The proof is
presented only for MPC without constraints. In
\cite{shi2007single}, the ILC update law is designed using MPC.
State constraints are not considered in \cite{shi2007single}. In
\cite{lee1999model} a batch MPC (BMPC) is proposed, which
integrates conventional MPC scheme with an iterative learning
scheme. A simplified static input-output map is considered in the
paper as opposed to a dynamical system.

In summary, we think that there is a need for more rigorous
theoretical justification attempted in this paper. Furthermore, to
the best of our knowledge, the literature on ILC-MPC schemes do
not consider state constraints, do not treat robust feasibility
issues in the MPC tracking problem, rigorous justification of
reference tracking proofs for the MPC is not present in the
literature and stability proofs for the combination of the ILC and
MPC schemes are not established in a systematic manner. Finally,
we want to cite the work of
\cite{aswani2013provably,aswani2012extensions,aswani2012reducing},
where similar control objectives as the one targeted in this
paper, have been studied using a learning-based MPC approach. The
main differences are in the control/learning design methodology
and the proof techniques.

{\it The main contribution of this work is to present a rigorous
proof of an ILC-MPC scheme using existing Lyapunov function based
stability analysis established in \cite{limon2010robust} and
extremum seeking algorithms in \cite{khong2013unified}, to justify
the modular design method for ILC-MPC proposed in \cite{Benosman},
where an ES-based modular approach to design ILC-MPC schemes for a
class of constrained linear systems is proposed.}

The rest of the paper is organized as follows. Section
\ref{sec_def_chap4} contains some useful notations and
definitions. The MPC control problem formulation is presented in
Section \ref{sec_pbformulation_chap4}. Section
\ref{direct-mpc-chapt4} is dedicated to a rigorous analysis of the
proposed ES-based ILC-MPC. Finally, some concluding comments are
presented in Section \ref{conclusion-chapt4}.

\section{Notation and basic definitions}\label{sec_def_chap4}
 Throughout this paper, $\Re$ denotes the set of real numbers and $\Ze$ denotes the set of
integers. State constraints and input constraints are represented
by $\X \subset \Re^n$ and $\U \subset \Re^m$ respectively. The
optimization horizon for MPC is denoted by $N \in \Ze_{\geq 1}$.
The feasible region for the MPC optimization problem is denoted by
$\X_N$. A continuous function $\alpha : \Re_{\geq 0} \imp
\Re_{\geq 0}$ with $\alpha(0)=0$ belongs to class $\mathcal{K}$ if
it is increasing and bounded. A function $\beta $ belongs to class
$\ki$ if it belongs to class $\mathcal{K}$ and is unbounded. A
function $\beta(s,t) \in \kl$ if $\beta(\cdot,t) \in \mathcal{K}$
and $\lim_{t \imp \infty}\beta(s,t)=0$. Given two sets $A$ and
$B$, such that $A\subset \Re^n$, $B\subset\Re^n$, the Minkowski
sum is defined as $A \oplus B:=\{a+b| a \in A, b \in B\}$. The
Pontryagin set difference is defined as $A \ominus B:= \{x| x
\oplus B \in A\}$. Given a matrix $M\in\Re^{m\times n}$, the set
$MA\subset\Re^m$, is defined as $MA\triangleq\{Ma\;:\;a\in A\}$. A
positive definite matrix is denoted by $P>0$. The standard
Euclidean norm is represented as $|x|$ for $x \in \Re^n$,
$|x|_{P}:=\sqrt{x^TPx}$ for a positive definite matrix $P$,
$|x|_{\mathcal{A}}:=\inf_{y \in \mathcal{A}}|x-y|$ for a closed
set $\mathcal{A} \subset \Re^n$ and $\|A\|$ represents an
appropriate matrix norm where $A$ is a matrix. $\Ball$ represents
the closed unit ball in the Euclidean space. Also, a matrix
$M\in\Re^{n\times n} $ is said to be Schur iff all its eigenvalues
are inside the unitary disk.

\section{Problem formulation}\label{sec_pbformulation_chap4}
In this section we will describe in detail the problem studied in
this paper. We consider linear systems of the form
\begin{eqnarray}\label{sys}
x(k+1)&=&(A+\Delta A)x(k)+(B+\Delta B)u(k),\\
y(k)&=&Cx(k)+Du(k),
\end{eqnarray}
where $\Delta A$ and $\Delta B$ represent the uncertainty in the
system model. We will assume that the uncertainties are bounded as
follows:

\begin{assumption}\label{asm:bounded}
The uncertainties $\|\Delta A\| \leq \ell_A$ and $\|\Delta B\|
\leq \ell_B$ for some $\ell_A,\ell_B > 0$.
\end{assumption}

Next, we impose some assumptions on the reference signal $r$.

\begin{assumption}\label{asm:ref}
The reference signal $r: [0,T] \imp \Re$ is a piecewise constant
trajectory for some $T>0$.
\end{assumption}

Under Assumptions \ref{asm:bounded} and \ref{asm:ref}, the goal is
to design a control scheme guarantying tracking with sufficiently
small errors by learning the uncertain parameters of the system.
Next, we will explain in detail the optimization problem
associated with the MPC based controller. The results stated here
are from \cite{limon2010robust}. We exploit the  analysis results
in \cite{limon2010robust} to establish that the closed-loop system
has an ISS property with respect to the parameter estimation
error.

Since the value of  $\Delta A$ and $\Delta B$ are not known a
priori, the MPC uses a model of the plant based on the current
estimate $\hat{\Delta}A $ and $\hat{\Delta}B$.

We will now formulate the MPC problem with a given estimate of the
uncertainty for a particular iteration of the learning process. We
will rewrite the system dynamics as
\begin{eqnarray}\label{sys}
x(k+1)=f(x,u)+g(x,u,\Delta)=F(x,u,\Delta),
\end{eqnarray}
where $f(x,u)=Ax+Bu$ and $g(x,u,\Delta)=\Delta A x + \Delta B u$.
\begin{assumption}\label{asm: State and control}
The state constraint set $\mathcal{X} \subset \Re^n$  and control
constraint set $\mathcal{U} \subset \Re^m $ are compact, convex
polyhedral sets.
\end{assumption}

The MPC model is generated using an estimate $\hat{\Delta} A $,
$\hat{\Delta} B $ and is expressed as
\begin{eqnarray}\label{sys}
x(k+1)=f(x,u)+g(x,u,\hat \Delta)=F(x,u,\hat\Delta).
\end{eqnarray}

We can now rewrite the actual model as
{\small\begin{eqnarray}\label{sys} x(k+1)=f(x,u)+g(x,u,\hat
\Delta)+(\Delta A-\hat \Delta A)x+(\Delta B-\hat \Delta B)u.
\end{eqnarray}}
This system can now be compared to the model in
\cite{limon2010robust}. So we have
\begin{eqnarray}
x(k+1)=F(x(k),u(k),\hat{\Delta})+w(k),
\end{eqnarray}
where
\begin{eqnarray}
w(k)=(\Delta A-\hat \Delta A)x(k)+(\Delta B-\hat \Delta B)u(k),
\end{eqnarray}
and $x(k) \in \mathcal{X}, u(k) \in \mathcal{U}$. The following
assumption will be justified in the next section.
\begin{assumption}\label{asm:esti}
The estimates of the uncertain parameters are bounded with $\|\hat
\Delta A\| \leq \ell_A$ and $\|\hat \Delta B\| \leq \ell_B$ for
all iterations of the extremum seeking algorithm.
\end{assumption}

We now impose certain conditions on the disturbance $w(k)$ and
system matrices in accordance with \cite[Assumption
1]{limon2010robust}.
\begin{assumption}
The pair $(A+\hat \Delta A, B + \hat \Delta B)$ is controllable
for every realization of $\hat \Delta A$ and $\hat \Delta B$.
\end{assumption}

We will denote the actual model using $(x,u)$ and the MPC model
through $(\bar{x},\bar{u})$. Hence we have
\begin{eqnarray*}
x(k+1)&=&F(x,u,\hat \Delta)+w,\\
\bar{x}(k+1)&=&F(\bar{x},\bar{u},\hat \Delta).
\end{eqnarray*}

\subsection{Robust positive invariant sets}
We denote the error between the states of the true model and MPC
model by $e(k)=x(k)-\bar{x}(k)$. We want the error to be bounded
during tracking. The error dynamics is then given by
\begin{eqnarray}\label{error_dynamic_uncertain_chapt4}
e(k+1)=(A+\hat \Delta A+(B+\hat \Delta B)K)e(k)+w(k),
\end{eqnarray}
where $u=\bar{u}+Ke$ and the matrix $K$ is such that $A_K:=(A+\hat
\Delta A+(B+\hat \Delta B)K)$ is Schur.

We first recall the definition of a robust positive invariant set
(RPI), e.g., \cite{limon2010robust}.
\begin{definition}
A set $\Phi_K$ is called an RPI set for the uncertain dynamics
(\ref{error_dynamic_uncertain_chapt4}), if $A_K\Phi_k\oplus
\mathcal{W}\subseteq \Phi_K$.
\end{definition}

So, we let $\Phi_K$ be an RPI set associated with the error
dynamics (\ref{error_dynamic_uncertain_chapt4}), i.e., $A_K \Phi_K
\oplus \mathcal{W} \subseteq \Phi_K$.

\subsection{Tightening the constraints}
Now we follow \cite{limon2010robust} and tighten the constraints
for the MPC model so that we achieve robust constraint
satisfaction for the actual model with uncertainties. Let $\X_1=X
\ominus \Phi_K$ and $\U_1 = \U \ominus K\Phi_K$. The following
result is from \cite[Proposition 1, Theorem 1 and Corollary 1
]{alvarado2007robust}.
\begin{proposition}
Let $\Phi_K$ be RPI for the error dynamics. If $e(0) \in \Phi_K$,
then $x(k) \in \bar{x}(k) \oplus \Phi_K$ for all $k \geq 0$ and
$w(k) \in \mathcal{W}$. If in addition, $\bar{x}(k) \in \X_1$ and
$\bar{u}(k) \in \U_1$ then with the control law $u=\bar{u}+Ke$,
$x(k) \in \X$ and $u(k) \in \U$ for all $k \geq 0$.
\end{proposition}

\subsection{Invariant set for tracking}
As in \cite{alvarado2007robust} and \cite{limon2010robust}, we
will characterize the set of nominal steady states and inputs so
that we can relate them later to the tracking problem. Let
$z_s=(\bar{x}_s,\bar{u}_s)$ be the steady state for the MPC model.
Then,
\begin{eqnarray}\label{steady}
\begin{bmatrix}
A+\hat \Delta A- I & B +\hat \Delta B\\
C & D
\end{bmatrix}
\begin{bmatrix}
\bar{x}_s\\
\bar{u}_s
\end{bmatrix}
=
\begin{bmatrix}
0\\
\bar{y}_s
\end{bmatrix}.
\end{eqnarray}
From the controllability assumption on the system matrices, the
admissible steady states can be characterized by a single
parameter $\bar{\theta}$ as
\begin{eqnarray}
\bar{z}_s&=& M_{\theta}\bar{\theta},\\
\bar{y}_s&=& N_{\theta}\bar{\theta},
\end{eqnarray}
for some $\bar{\theta}$ and matrices $M_{\theta}$ and
$N_{\theta}=[C\;\; D]M_{\theta}$. We let $\X_s$, $\U_s$ denote the
set of admissible steady states that are contained in $\X_1,\U_1$
and satisfy \eqref{steady}. $\mathcal{Y}_s$ denotes the set of
admissible output steady states. Now we will define an invariant
set for tracking which will be utilized as a terminal constraint
for the optimization problem.

\begin{definition}\cite[Definition 2]{limon2010robust}
An invariant set for tracking for the MPC model is the set of
initial conditions, steady states and inputs (characterized by
$\bar{\theta}$) that can be stabilized by the control law
$\bar{u}=\bar{K}\bar{x}+L\bar{\theta}$ with $L:=[-\bar{K}\;\;
I]M_{\theta}$ while $(\bar{x}(k),\bar{u}(k)) \in \X_1 \times \U_1$
for all $k \geq 0$.
\end{definition}

We choose the matrix $\bar{K}$ such that $A_{\bar{K}}:=(A+\hat
\Delta A+(B+\hat \Delta B)\bar{K})$ is Schur. We refer the reader
to \cite{alvarado2007robust} and \cite{limon2010robust} for more
details on computing the invariant set for tracking. We will refer
to the invariant set for tracking as $\Omega_{\bar{K}}$. We say a
point $(\bar{x}(0),\bar{\theta}) \in \Omega_{\bar{K}}$ if with the
control law $u=\bar{K}(\bar{x}-\bar{x}_s)+
\bar{u}_s=\bar{K}\bar{x}+L\bar{\theta}$, the solutions of the MPC
model from $\bar{x}(0)$ satisfy
$\bar{x}(k) \in \text{Proj}_x(\Omega_{\bar{K}})$ for all $k \geq
0$. As stated in \cite{limon2010robust} the set can be taken to be
a polyhedral.

\subsection{MPC Optimization problem}
Now we will define the optimization problem that will be solved at
every instant to determine the control law for the actual plant
dynamics. For a given target setpoint $y_t$ and initial condition
$x$, the  optimization problem $\mathcal{P}_N(x,y_t)$ is defined
as,
\begin{align*}
&\min_{\bar{x}(0),\bar{\theta},\bar{\textbf{u}}} V_N(x,y_t,\bar{x}(0),\bar{\theta},\bar{\textbf{u}})\\
&\text{s.t}\;\; \bar{x}(0) \in x \oplus (-\Phi_K)\\
& \bar{x}(k+1)= (A+\hat \Delta A)\bar{x}(k)+(B+\hat \Delta B)\bar{u}(k)\\
& \bar{x}_s = M_{\theta}\bar{\theta}\\
& \bar{y}_s = N_{\theta}\bar{\theta}\\
& (\bar{x}(k),\bar{u}(k)) \in \X_1 \times \U_1, k \in \Ze_{\leq N-1} \\
& (\bar{x}(N),\bar{\theta}) \in \Omega_{\bar{K}},
\end{align*}
where the cost function is defined as follows
\begin{eqnarray}
\nonumber V_N(x,y_t,\bar{x}(0),\bar{\theta},\bar{\textbf{u}})=\sum_{k=0}^{N-1}|\bar{x}(k)-\bar{x}_s|_{\tilde{Q}} ^2\\
+|\bar{u}(k)-\bar{u}_s|_R^2+ |\bar{x}(N)-\bar{x}_s|_P^2+
|\bar{y}_s-y_t|_T^2.
\end{eqnarray}
Such cost function is frequently used in MPC literature for
tracking except for the additional term in the end which penalizes
the difference between the artificial stead state and the actual
target value. We refer the reader to \cite{alvarado2007output},
\cite{alvarado2007robust} and \cite{limon2010robust} for more
details.

\begin{assumption}
The following conditions are satisfied by the optimization problem
\begin{enumerate}
\item The matrices $\tilde{Q}>0, R>0, T> 0$. \item $(A+\hat \Delta
A+ (B+\hat \Delta B)K)$ is Schur matrix, $\Phi_K$ is a RPI set for
the error dynamics, and $\X_1,\U_1$ are non-empty. \item The
matrix $\bar{K}$ is such that $A+\hat \Delta A+(B+\hat \Delta
B)\bar{K}$ is Schur and $P>0$ satisfies: {\small\begin{eqnarray*}
\hspace{-0.5cm}P-(A+\hat \Delta A+(B+\hat \Delta B)\bar{K})^TP
\hspace{0mm} (A+\hat \Delta A+(B+\hat \Delta B)\bar{K})
=\\\tilde{Q}+\bar{K}^TR\bar{K}.
\end{eqnarray*}}
\item The set $\Omega_{\bar{K}}$ is an invariant set for tracking
subject to  the tightened constraints $\X_1,\U_1$.
\end{enumerate}
\end{assumption}

As noted in \cite{limon2010robust}, the  feasible set $\X_N$ does
not vary with the set points $y_t$ and the optimization problem is
a Quadratic programming (QP) problem. The optimal values are given
by $\bar{x}_s^*,\bar{u}^*(0), \bar{x}^*$. The MPC control law
writes then as: $u=\kappa_N(x)=K(x-\bar{x}^*)+\bar{u}^*(0)$. The
MPC law $\kappa_N$ implicitly depends on the current estimate of
the uncertainty $\hat \Delta$. Also it follows from the results in
\cite{bemporad2002explicit} that the control law for the MPC
problem is continuous\footnote{The authors would like to thank Dr.
S. Di Cairano for pointing out to us the paper
\cite{bemporad2002explicit}.}.

\section{DIRECT extremum seeking-based iterative learning MPC}\label{direct-mpc-chapt4}
\subsection{DIRECT-based iterative learning MPC}
In this section we will explain the assumptions regarding the
learning cost function\footnote{Not to be confused with the MPC
cost function.} used for identifying the true parameters of the
uncertain system via nonlinear programming based extremum seeking
called DIRECT, e.g., \cite{jones1993lipschitzian}. Let $\Delta$ be
a vector that contains the entries in $\Delta A $ and $\Delta B$.
Similarly the estimate will be denoted by $\hat \Delta$. Then
$\Delta,\hat{\Delta} \in \Re^{n(n+m)}$.

Since we do not impose the presence of attractors for the
closed-loop system as in \cite{popovic2006extremum} or
\cite{khong2013multidimensional}, the cost function that we
utilize $Q: \Re^{n(n+m)} \imp \Re_{\geq 0}$ depends on $x_0$. For
iterative learning methods, the same initial condition $x_0$ is
used to learn the uncertain parameters and hence we refer to
$Q(x_0,\hat \Delta)$ as only $Q(\hat \Delta)$ since $x_0$ is
fixed.

\begin{assumption}\label{di2}
The learning cost function $Q: \Re^{n(n+m)} \imp \Re_{\geq 0}$ is
\begin{enumerate}
\item Lipshitz in the compact set of uncertain parameters \item
The true parameter $ \Delta$ is such that $Q(\Delta) < Q(\hat
\Delta)$ for all $\hat \Delta \neq \Delta$.
\end{enumerate}
\end{assumption}

One example of a learning cost functions is identification-type
cost function, where the error between outputs measurements from
the system are compared to the MPC model outputs. Another example
of a learning cost function, can be a performance-type cost
function, where a measured output of the system is directly
compared to a desired reference trajectory.

We then use the DIRECT optimization algorithm introduced in
\cite{jones1993lipschitzian} for finding the global minimum of a
Lipschitz function without knowledge of the Lipschitz constant.
The algorithm is implemented in MATLAB using
\cite{finkel2003direct}. We will utilize a modified termination
criterion introduced in \cite{khong2013multidimensional} for the
DIRECT algorithm to make it more suitable for extremum seeking
applications. As we will mention in later sections, the DIRECT
algorithm has nice convergence properties which will be used to
establish our main results.

\subsection{Main results: Proof of the MPC ISS and the learning convergence}
We will now present the main results of this paper, namely the
stability analysis of the proposed ILC-MPC algorithm, using the
existing results for MPC tracking and DIRECT algorithm established
in \cite{limon2010robust} and \cite{khong2013unified},
respectively.

First, we define the value function
$$V_N^*(x,y_t)=\min_{\bar{x}(0),\theta,\bar{\textbf{u}}}V_N(x,y_t,\bar{x}(0),\theta,\bar{\textbf{u}})$$
for a fixed target $y_t$. Also, we let $\tilde
\theta:=\arg\min_{\bar{\theta}}|N_{\theta}\bar{\theta}-y_t|$,
$(\tilde{x}_s, \tilde{u}_s)=M_{\theta}\tilde{\theta}$ and
$\tilde{y}_s=C\tilde{x}_s+D\tilde{u}_s$. If the target steady
state $y_t$ is not admissible, the MPC tracking scheme drives the
output to converge to the point $\tilde{y}_s$ which is a steady
state output that is admissible and also minimizes the error with
the target steady state, i.e., graceful target degradation
principle, e.g., \cite{BL09-2}. The proof of the following result
follows  from \cite[Theorem 1]{limon2010robust} and classically
uses $V_N^*(x,y_t)$ as a Lyapunov function for the closed-loop
system.

\begin{proposition}\label{main 1}
Let $y_t$ be given. For all $x(0) \in \mathcal{X}_N$, the MPC
problem is recursively feasible. The state $x(k)$ converges to
$\tilde{x}_s \oplus \Phi_K$ and the output $y(k)$ converges to
$\tilde{y}_s \oplus (C+DK)\Phi_K$.
\end{proposition}

The next result states the convergence properties of the modified
DIRECT algorithm, which we will used in establishing the main
result. This result is stated as \cite[Assumption
7]{khong2013unified} and it follows from the analysis of the
modified DIRECT algorithm in \cite{khong2013multidimensional}.
\begin{proposition}\label{main 2}

For any sequence of updates $\hat{\Delta}_{t},\;t=1,2,...$ from
the modified DIRECT algorithm and $\varepsilon > 0$, there exists
a $N> 0$ such that $|\Delta- \hat{\Delta}_t| \leq \varepsilon $
for $t \geq N$.
\end{proposition}

\begin{remark}
Note that the results in \cite{khong2013multidimensional} also
include a robustness aspect  of the DIRECT algorithm. This can be
used to account for measurement noises and computational error
associated with the learning cost $Q$.
\end{remark}

We now state the main result of the section that combines the ISS
MPC formulation and the extremum seeking algorithm.

\begin{theorem}
Under Assumptions 1-7, given an initial condition $x_0$, an output
target $y_t$, such that $y_t$ is constant over $[0,T*]$ for some
$T*$ sufficiently large. Then, for every $\varepsilon>0$, there
exists $N_{1}$ and $N_{2}$ such that $|y(k)-\tilde{y}_s| <=
\varepsilon$ for $k \in [N_1, T*]$ after $N_2$ iterations  of the
ILC-MPC scheme.

\end{theorem}
\begin{pf}
It can observed that since the size of $\Phi_K$ grows with the
size of $\mathcal{W}$ and $\Phi_K=\{0\}$ for the case without
disturbances that without loss of generality $\Phi_K \subseteq
\Gamma(w^*)\Ball$, where $\Gamma \in \mathcal{K}$ and  $w^*=
\|\Delta A- \hat{\Delta}A\|X^*+ \|\Delta B - \hat{\Delta}B\|U^*$,
where $X^* = \max_{x \in \X}|x|$ and $U^*=\max_{u \in \U}|u|$.
Here $X^*,U^*$ are fixed over both regular time and learning
iteration number, but the uncertainties vary over iterations
because of the modified DIRECT algorithm updates. Since the worst
case disturbance depends directly on the estimation error, without
loss of generality we have that $\Phi_K \subseteq \gamma(|\Delta -
\hat{\Delta}|)\Ball$ and $(C+DK)\Phi_K \subseteq
{\gamma^*}(|\Delta - \hat \Delta|)\Ball$ for some $\gamma,\gamma^*
\in \mathcal{K}$ . It follows from Proposition \ref{main 1} that
$\lim_{k \imp \infty}|x(k)|_{\tilde{x}_s \oplus \Phi_K}=0.$  Then,
 \begin{eqnarray*}
 \lim_{k \imp \infty}|x(k)-\tilde{x}_s| &\leq& \max_{x \in \Phi_K}|x|\\
&\leq& \gamma(|\Delta- \hat \Delta|).
\end{eqnarray*}
We observe that the above set of equations state that the
closed-loop system with the MPC controller has  the asymptotic
gain property and it is upper bounded by the size of the parameter
estimation error. Note that the estimate $\hat \Delta$ is constant
for a particular iteration of the process. Also, for the case of
no uncertainties we have $0-$stability (Lyapunov stability for the
case of zero uncertainty). This can be proven by using the cost
function $V_N^*(x,y_t)$ as the Lyapunov function, such that
$V_N^*(x(k+1),y_t) \leq V_N^*(x(k),y_t)$ and
$\lambda_{\min}(\tilde{Q})|x-\tilde{x}_s|^2 \leq V_N^*(x,y_t) \leq
\lambda_{\max}(P)|x-\tilde{x}_s|^2$, see \cite{limon2008mpc}.
Furthermore, here the stability and asymptotic gain property can
be interpreted with respect to the compact set
$\mathcal{A}:=\{\tilde{x}_s\}$.

Since the MPC control law is continuous, the closed-loop system
for a particular iteration of the ILC-MPC scheme is also
continuous with respect to the state. Then, from \cite[Theorem
3.1]{cai2009characterizations}  we can conclude that the
closed-loop system is ISS with respect to the parameter estimation
error and hence satisfies,
\begin{eqnarray*}
|x(k)-\tilde{x}_s| \leq \beta(|x(0)-\tilde{x}_s|,k)+
\hat{\gamma}(|\Delta-\hat \Delta|),
\end{eqnarray*}

where $\beta \in \kl$ and $\hat{\gamma} \in \mathcal{K}$. Now, let
$\varepsilon_1 > 0$ be small enough such that
$\hat\gamma(\varepsilon_1) \leq \varepsilon/2$. From Proposition
\ref{main 2}, it follows that there exists a $N_2>0$ such that
$|\Delta - \hat{\Delta}_t| \leq \varepsilon_1$ for $t \geq N_2$,
where $t$ is the iteration number of the ILC-MPC scheme. Hence
there exists $N_1 > 0$ such that
$|\beta(|x(0)-\tilde{x}_s|,k)|\leq \varepsilon/2$ for $k \geq
N_1$. We choose $T^*$ such that $T^*> N_1$. Then, we have that for
$k \in [N_1,T^*]$ and for $t \geq N_2$,
\begin{eqnarray*}
|x(k)-\tilde{x}_s| &\leq& \varepsilon.
\end{eqnarray*}

Similarly, using the linearity dependence between $y$ and $x$, we
can also establish that,
$\exists\;\tilde\varepsilon(\varepsilon)$, such that for $k \in
[N_1,T^*]$ and for $t \geq N_2$
\begin{eqnarray*}
|y(k)-\tilde{y}_s| &\leq& \tilde\varepsilon(\varepsilon).
\end{eqnarray*}
\end{pf}

\section{Conclusion}\label{conclusion-chapt4}

In this paper, we have reported some results about extremum
seeking-based ILC-MPC algorithms. We have argued that it is
possible to merge together a model-based linear MPC algorithm with
a model-free ES algorithm to iteratively learn structural model
uncertainties and thus improve the overall performance of the MPC
controller. We have presented the stability analysis of this
modular design technique for ES-based ILC-MPC. where we addressed
both feasibility and tracking performance. Future work can include
extending this method to a wider class of nonlinear systems,
tracking a more richer class of signals, employing different
non-smooth optimization techniques for the extremum seeking
algorithm, etc.

\bibliographystyle{IEEEtran}

\bibliography{C:/benosman/adaptive_inovative_control/book_chapter/Book_chapters/Chapter_one/bibliov3}

\begin{thebibliography}{10}
\providecommand{\url}[1]{#1}
\def\UrlFont{\rmfamily}
\providecommand{\newblock}{\relax}
\providecommand{\bibinfo}[2]{#2}
\providecommand\BIBentrySTDinterwordspacing{\spaceskip=0pt\relax}
\providecommand\BIBentryALTinterwordstretchfactor{4}
\providecommand\BIBentryALTinterwordspacing{\spaceskip=\fontdimen2\font plus
\BIBentryALTinterwordstretchfactor\fontdimen3\font minus
  \fontdimen4\font\relax}
\providecommand\BIBforeignlanguage[2]{{%
\expandafter\ifx\csname l@#1\endcsname\relax
\typeout{** WARNING: IEEEtran.bst: No hyphenation pattern has been}%
\typeout{** loaded for the language `#1'. Using the pattern for}%
\typeout{** the default language instead.}%
\else
\language=\csname l@#1\endcsname
\fi
#2}}

\bibitem{B014}
M.~Benosman, ``Learning-based adaptive control for nonlinear systems,'' in
  \emph{European Control Conference}, 2014, pp. 920--925.

\bibitem{Mayne2000}
D.~Q. Mayne, J.~B. Rawlings, C.~V. Rao, and P.~O.~M. Scokaert, ``Constrained
  model predictive control: Stability and optimality,'' \emph{Automatica},
  vol.~36, pp. 789--814, 2000.

\bibitem{ariyur2003real}
K.~B. Ariyur and M.~Krstic, \emph{Real-time optimization by extremum-seeking
  control}.\hskip 1em plus 0.5em minus 0.4em\relax John Wiley \& Sons, 2003.

\bibitem{AK02}
------, ``Multivariable extremum seeking feedback: Analysis and design,'' in
  \emph{Proc. of the Mathematical Theory of Networks and Systems}, South Bend,
  IN, August 2002.

\bibitem{Nes09}
D.~Ne{s}ic, ``Extremum seeking control: Convergence analysis,'' \emph{European
  Journal of Control}, vol.~15, no. 3–4, pp. 331 -- 347, 2009.

\bibitem{k00}
M.~Krstic, ``Performance improvement and limitations in extremum seeking,''
  \emph{Systems $\&$ Control Letters}, vol.~39, pp. 313--326, 2000.

\bibitem{TNM06}
Y.~Tan, D.~Nesic, and I.~Mareels, ``On non-local stability properties of
  extremum seeking control,'' \emph{Automatica}, no.~42, pp. 889--903, 2006.

\bibitem{rotea2000analysis}
M.~Rotea, ``Analysis of multivariable extremum seeking algorithms,'' in
  \emph{Proceedings of the American Control Conference}, vol.~1, no.~6.\hskip
  1em plus 0.5em minus 0.4em\relax IEEE, 2000, pp. 433--437.

\bibitem{GDD13}
M.~Guay, S.~Dhaliwal, and D.~Dochain, ``A time-varying extremum-seeking control
  approach,'' in \emph{American Control Conference}, 2013.

\bibitem{ZGD03}
T.~Zhang, M.~Guay, and D.~Dochain, ``Adaptive extremum seeking control of
  continuous stirred-tank bioreactors,'' \emph{AIChE J.}, no.~49, p. 113–123.,
  2003.

\bibitem{HGPD08}
N.~Hudon, M.~Guay, M.~Perrier, and D.~Dochain, ``Adaptive extremum-seeking
  control of convection-reaction distributed reactor with limited actuation,''
  \emph{Computers $\&$ Chemical Engineering}, vol.~32, no.~12, pp. 2994 --
  3001, 2008.

\bibitem{ZO12}
C.~Zhang and R.~Ordóñez, \emph{Extremum-Seeking Control and
  Applications}.\hskip 1em plus 0.5em minus 0.4em\relax Springer-Verlag, 2012.

\bibitem{benosmanatincacc13}
M.~Benosman and G.~Atinc, ``Multi-parametric extremum seeking-based learning
  control for electromagnetic actuators,'' in \emph{American Control
  Conference}, 2013.

\bibitem{benosmanatincecc13}
------, ``Nonlinear learning-based adaptive control for electromagnetic
  actuators,'' in \emph{European Control Conference}, 2013.

\bibitem{benosmanatinccdc13}
G.~Atinc and M.~Benosman, ``Nonlinear learning-based adaptive control for
  electromagnetic actuators with proof of stability,'' in \emph{IEEE,
  Conference on Decision and Control}, 2013.

\bibitem{arimoto1990robustness}
S.~Arimoto, ``Robustness of learning control for robot manipulators,'' in
  \emph{Proceedings of the IEEE International Conference on Robotics and
  Automation}.\hskip 1em plus 0.5em minus 0.4em\relax IEEE, 1990, pp.
  1528--1533.

\bibitem{wang2009survey}
Y.~Wang, F.~Gao, and F.~J. Doyle~III, ``Survey on iterative learning control,
  repetitive control, and run-to-run control,'' \emph{Journal of Process
  Control}, vol.~19, no.~10, pp. 1589--1600, 2009.

\bibitem{moore1999iterative}
K.~L. Moore, ``Iterative learning control: an expository overview,'' in
  \emph{Applied and computational control, signals, and circuits}.\hskip 1em
  plus 0.5em minus 0.4em\relax Springer, 1999, pp. 151--214.

\bibitem{ahn2007iterative}
H.-S. Ahn, Y.~Chen, and K.~L. Moore, ``Iterative learning control: brief survey
  and categorization,'' \emph{IEEE Transactions on Systems Man and
  Cybernetics}, vol.~37, no.~6, p. 1099, 2007.

\bibitem{wang2008iterative}
Y.~Wang, D.~Zhou, and F.~Gao, ``Iterative learning model predictive control for
  multi-phase batch processes,'' \emph{Journal of Process Control}, vol.~18,
  no.~6, pp. 543--557, 2008.

\bibitem{cueli2008iterative}
J.~R. Cueli and C.~Bordons, ``Iterative nonlinear model predictive control.
  stability, robustness and applications,'' \emph{Control Engineering
  Practice}, vol.~16, no.~9, pp. 1023--1034, 2008.

\bibitem{shi2007single}
J.~Shi, F.~Gao, and T.-J. Wu, ``Single-cycle and multi-cycle generalized 2{D}
  model predictive iterative learning control (2{D}-{GPILC}) schemes for batch
  processes,'' \emph{Journal of Process Control}, vol.~17, no.~9, pp. 715--727,
  2007.

\bibitem{lee1999model}
K.~S. Lee, I.-S. Chin, H.~J. Lee, and J.~H. Lee, ``Model predictive control
  technique combined with iterative learning for batch processes,'' \emph{AIChE
  Journal}, vol.~45, no.~10, pp. 2175--2187, 1999.

\bibitem{aswani2013provably}
A.~Aswani, H.~Gonzalez, S.~S. Sastry, and C.~Tomlin, ``Provably safe and robust
  learning-based model predictive control,'' \emph{Automatica}, vol.~49, no.~5,
  pp. 1216--1226, 2013.

\bibitem{aswani2012extensions}
A.~Aswani, P.~Bouffard, and C.~Tomlin, ``Extensions of learning-based model
  predictive control for real-time application to a quadrotor helicopter,'' in
  \emph{American Control Conference (ACC), 2012}.\hskip 1em plus 0.5em minus
  0.4em\relax IEEE, 2012, pp. 4661--4666.

\bibitem{aswani2012reducing}
A.~Aswani, N.~Master, J.~Taneja, D.~Culler, and C.~Tomlin, ``Reducing transient
  and steady state electricity consumption in hvac using learning-based
  model-predictive control,'' \emph{Proceedings of the IEEE}, vol. 100, no.~1,
  pp. 240--253, 2012.

\bibitem{limon2010robust}
D.~Limon, I.~Alvarado, T.~Alamo, and E.~Camacho, ``Robust tube-based {MPC} for
  tracking of constrained linear systems with additive disturbances,''
  \emph{Journal of Process Control}, vol.~20, no.~3, pp. 248--260, 2010.

\bibitem{khong2013unified}
S.~Z. Khong, D.~Ne{\v{s}}i{\'c}, Y.~Tan, and C.~Manzie, ``Unified frameworks
  for sampled-data extremum seeking control: Global optimisation and multi-unit
  systems,'' \emph{Automatica}, vol.~49, no.~9, pp. 2720--2733, 2013.

\bibitem{Benosman}
M.~Benosman, S.~D. Cairano, and A.~Weiss, ``Extremum seeking-based iterative
  learning linear {MPC},'' in \emph{IEEE Multi-conference on Systems and
  Control}, 2014.

\bibitem{alvarado2007robust}
I.~Alvarado, D.~Limon, T.~Alamo, M.~Fiacchini, and E.~Camacho, ``Robust tube
  based {MPC} for tracking of piece-wise constant references,'' in
  \emph{Decision and Control, 2007 46th IEEE Conference on}.\hskip 1em plus
  0.5em minus 0.4em\relax IEEE, 2007, pp. 1820--1825.

\bibitem{alvarado2007output}
I.~Alvarado, D.~Limon, T.~Alamo, and E.~Camacho, ``Output feedback robust tube
  based {MPC} for tracking of piece-wise constant references,'' in
  \emph{Proceedings of the 46th IEEE Conference on Decision and Control}.\hskip
  1em plus 0.5em minus 0.4em\relax IEEE, 2007, pp. 2175--2180.

\bibitem{bemporad2002explicit}
A.~Bemporad, M.~Morari, V.~Dua, and E.~N. Pistikopoulos, ``The explicit linear
  quadratic regulator for constrained systems,'' \emph{Automatica}, vol.~38,
  no.~1, pp. 3--20, 2002.

\bibitem{jones1993lipschitzian}
D.~R. Jones, C.~D. Perttunen, and B.~E. Stuckman, ``Lipschitzian optimization
  without the {L}ipschitz constant,'' \emph{Journal of Optimization Theory and
  Applications}, vol.~79, no.~1, pp. 157--181, 1993.

\bibitem{popovic2006extremum}
D.~Popovic, M.~Jankovic, S.~Magner, and A.~R. Teel, ``Extremum seeking methods
  for optimization of variable cam timing engine operation,'' \emph{IEEE
  Transactions on Control Systems Technology}, vol.~14, no.~3, pp. 398--407,
  2006.

\bibitem{khong2013multidimensional}
S.~Z. Khong, D.~Ne{\v{s}}i{\'c}, C.~Manzie, and Y.~Tan, ``Multidimensional
  global extremum seeking via the {DIRECT} optimisation algorithm,''
  \emph{Automatica}, vol.~49, no.~7, pp. 1970--1978, 2013.

\bibitem{finkel2003direct}
D.~E. Finkel, ``{DIRECT} optimization algorithm user guide,'' \emph{Center for
  Research in Scientific Computation, North Carolina State University}, vol.~2,
  2003.

\bibitem{BL09-2}
M.~Benosman and K.-Y. Lum, ``On-line references reshaping and control
  re-allocation for nonlinear fault tolerant control,'' \emph{IEEE,
  Transactions on Control Systems Technology}, vol.~17, no.~2, pp. 366--379,
  March 2009.

\bibitem{limon2008mpc}
D.~Lim{\'o}n, I.~Alvarado, T.~Alamo, and E.~F. Camacho, ``{MPC} for tracking
  piecewise constant references for constrained linear systems,''
  \emph{Automatica}, vol.~44, no.~9, pp. 2382--2387, 2008.

\bibitem{cai2009characterizations}
C.~Cai and A.~R. Teel, ``Characterizations of input-to-state stability for
  hybrid systems,'' \emph{Systems \& Control Letters}, vol.~58, no.~1, pp.
  47--53, 2009.

\end{thebibliography}

\end{document}